\documentclass[11pt,a4paper]{JHEP3}
\usepackage{amsmath, amssymb}
\usepackage{euscript}
\def\be{\begin{eqnarray}}
\def\ee{\end{eqnarray}}
\def\0{\nonumber}
\def\d{\partial}

\usepackage{amsfonts}
\usepackage{verbatim}
\newcommand{\CR}{\\\nonumber}

\newcommand{\refb}[1]{(\ref{#1})}
\newcommand{\NO}[1]{\ensuremath{\,:\!#1\!:\,}}
\newcommand{\VEV}[1]{\ensuremath{\,\langle{}#1\rangle\,}}

\newcommand{\Tr}{\mathrm{Tr}}

\newcommand\la{\lambda}

\def\k{\kappa}
\def\u{\tilde u}

\preprint{SISSA/15/2008/EP\\\tt hep-th/0804.0198}

\title{Hawking radiation, $W_\infty$ algebra and trace anomalies}

\author{ L.Bonora$^a$, M.Cvitan$^{a,b}$ \\
 $^a$ International School for Advanced Studies (SISSA/ISAS)\\
Via Beirut 2--4, 34014 Trieste, Italy, and INFN, Sezione di
Trieste\\
  $^b$ Theoretical Physics Department, Faculty of Science,
        University of Zagreb\\
        p.p. 331, HR-10002 Zagreb, Croatia\\

E-mail:   \email{bonora@sissa.it}, \email{cvitan@sissa.it}}

\abstract{We apply the ``trace anomaly method'' to the calculation of 
moments of the Hawking radiation of a Schwarzschild black hole. 
We show that they can be explained as the fluxes of chiral
currents forming a $W_\infty$ algebra.
Then we construct the covariant version of these 
currents and verify that up to order 6 they are not affected by any 
trace anomaly. Using cohomological methods we show that actually, for 
the fourth order current, no trace anomalies can exist. The  
results reported here are strictly valid in two dimensions.}


\begin{document}

\maketitle

\section{Introduction}

In the last few years there has been an increasing activity in 
calculating the Hawking radiation \cite{Hawking1,Hawking2} 
by means of anomalies. This 
renewed attention to the relation between anomalies and Hawking
radiation was pioneered by the paper \cite{Robinson}, which was 
followed by several other contributions \cite{IUW1,IUW2,IMU1,IMU2,IMU3,
IMU4,IMU5,
Murata:2006pt,
Vagenas:2006qb,
Setare:2006hq,
Jiang:2007gc,
Jiang:2007pn,
Jiang:2007wj,
Kui:2007dy,
Shin:2007gz,
Jiang:2007mi,
Das:2007ru,
Chen:2007pp,
Miyamoto:2007ue,
Jiang:2007pe,
Kim:2007ge,
Murata:2007zr,
Peng:2007nj,
Ma:2007xr,
Huang:2007ed,
Peng:2008ru,
Wu:2008yx,
Gangopadhyay:2008zw,
Kim:2008hm,
Xu,Banerjee1,Banerjee2,Gango,Gango2,Kulkarni,Peng1,Peng2,Peng3}. 
In \cite{Robinson} the method used was based on the 
diffeomorphism anomaly in a two--dimensional effective field theory
near the horizon of a radially symmetric static black hole.
The argument is that, since just outside the horizon only outgoing 
modes may exist, the physics near the horizon can be described by 
an effective two--dimensional chiral field theory
(of infinite many fields) in which ingoing modes
have been integrated out. This implies 
an effective breakdown of the 
diff invariance. The ensuing anomaly equation can be utilized to compute
the outgoing flux of radiation. 

A different method, based on trace anomaly, had been suggested long ago
by Christensen and Fulling, \cite{CF}. This method provides a full 
solution only in two dimensions, the reason being that its utilization 
involves the region away from the horizon, where a two-dimensional
formalism does not provide a good description. The method 
has been reproposed in different forms in \cite{Thorlacius,Strominger} 
and, in particular, 
\cite{IMU2} and \cite{IMU4}. In this paper we would like to discuss
a few aspects of the trace anomaly method and its implications. In 
\cite{IMU2} the authors made the remarkable observation that the full 
spectrum of the Planck distribution of a thermal Hawking radiation
of a Schwarzschild black hole can be described by postulating the
existence, in the two--dimensional effective field theory near the horizon,
of higher spin currents and applying a generalization of the
trace anomaly method. These authors in subsequent papers fully developed
this method for fermionic currents. In this paper we do the same 
for bosonic higher spin currents. This allows us to clarify, first of all,
that the higher spin currents necessary to reproduce the thermal
Hawking radiation form a $W_\infty$ algebra. We then covariantize
the higher spin currents, according to the method proposed in \cite{IMU4},
but, differently the latter reference, we do not find any trace 
anomaly in the higher spin currents. This prompts us to analyze the nature
of these anomalies. Using consistency methods we find that the trace anomalies
of ref.\cite{IMU4} are cohomologically trivial. This means that they are
an artifact of the regularization employed.

\section{$W_\infty$ algebra and Hawking radiation}

Let us review the argument that allows us to evaluate the outgoing 
radiation from a Schwarzschild black hole starting from the 
trace anomaly of the energy--momentum tensor 
(we closely follow \cite{IMU2}). Here we assume the point of 
view, advocated by several authors \cite{Thorlacius,Strominger} and in particular in 
\cite{Robinson}, that near--horizon physics is described by a 
two--dimensional conformal field theory (see also \cite{Carlip,Solod}). 
Due to the Einstein
equation, the trace of the matter energy momentum tensor vanishes on shell.
However it is generally the case that the latter is nonvanishing at one 
loop, due to an anomaly: $T_\alpha^\alpha= \frac c{24\pi} R$ where 
$R$ is the background Ricci scalar. $c$ is the central charge of the 
matter system. This is no accident, in fact it is well--known that the
above trace anomaly is related to the cocycle that pops up in the
conformal transformation of the (holomorphic or anti--holomorphic part
of the) energy momentum tensor. If the matter system is chiral, this 
cocycle also determines the diffeomorphism anomaly (which we do not 
consider in this paper). 

In light--cone coordinates $u=t-r_*, v=t+r_*$, let us denote
by $T_{uu}(u,v)$ and $T_{vv}(u,v)$ the classically non vanishing
components of the energy--momentum tensor. Given a background metric
$g_{\alpha\beta} = e^\varphi \eta_{\alpha\beta}$, the trace anomaly equation (together with the conservation equation)
can be solved. It yields
\be
T_{uu}(u,v)= \frac c{24\pi} \left(\d_u^2 \varphi- 
\frac 12 (\d_u\varphi)^2\right)+T^{(hol)}_{uu} (u)\label{Tuu}
\ee
where $T_{uu}^{(hol)}$ is holomorphic, while $T_{uu}$ is conformally 
covariant. I.e., under a conformal transformation $u\to \tilde u=f(u)
(v\to \tilde v=g(u))$ one has
\be
T_{uu}(u,v)=\left(\frac {df}{du}\right)^2 T_{\tilde u\tilde u}
(\tilde u, \tilde v)\label{conf}  
\ee 
Since, under a conformal transformation, $\tilde \varphi(\u,\tilde v)=
\varphi(u,v) -\ln \left( \frac {df}{du}\frac {dg}{dv}\right)$, it follows 
that 
\be
T^{(hol)}_{\u \u}(\u) = \left(\frac {df}{du} \right)^{-2} \left(
T^{(hol)}_{u u}(u)+ \frac c{24\pi} \{\u,u\}\right)\label{Thol}
\ee
Regular coordinates near the horizon are the Kruskal ones, $(U,V)$, defined by
$U=-e^{-\k u}$ and $V=e^{\k v}$. Under this transformation we have
\be
T^{(hol)}_{UU}(U) = \left(\frac 1{\k U} \right)^{2} \left(
T^{(hol)}_{u u}(u)+ \frac c{24\pi} \{U,u\}\right)\label{TholU}
\ee
Now we require the outgoing energy flux to be regular at the 
horizon $U=0$ in the Kruskal coordinate. Therefore at that point 
$T^{(hol)}_{u u}(u)$ is given by $\frac{c \k^2}{48 \pi}$. 
Since the background is static, $T^{(hol)}_{u u}(u)$
is constant in $t$ and therefore also in $r$. Therefore
$\frac{c \k^2}{48 \pi}$ is its value also at $r=\infty$.
On the other hand we can assume that at $r=\infty$ 
there be no incoming flux and that the background be 
trivial  (so that the vev of $T^{(hol)}_{u u}(u)$ and $T_{u u}(u,v)$ 
asymptotically coincide). 
Therefore the asymptotic flux is 
\be
\langle T_t^r\rangle =\VEV {T_{uu}}-\VEV {T_{vv}} = \frac{c \k^2}{48 \pi}  
\label{flux}
\ee

Now let the thermal bosonic spectrum of the black hole, due to 
emission of a scalar complex boson ($c=2)$, be given by the 
Planck distribution
\be
N(\omega)= \frac 2{e^{\beta\omega}-1} \label{Planck}
\ee
where $1/\beta$ is the Hawking temperature and $\omega=|k|$. 
In two dimensions the flux moments are defined by
\be
F_n= \frac 1{4\pi} \int_{-\infty}^{+\infty} dk \frac {\omega\,k^{n-2} }
{e^{\beta\omega}-1} \0
\ee 
They vanish for $n$ odd, while for $n$ even they are given by
\be
F_{2n}=\frac 1{4\pi}\int_0^\infty d\omega \omega^{2n-1} N(\omega)=
\frac {2(-1)^{n+1}}{8\pi n} B_{2n} \k^{2n}\label{moments} 
\ee
where $B_n$ are the Bernoulli numbers 
($B_2=\frac 16,B_4=-\frac 1{30},.. $).
Therefore 
the outgoing flux (\ref{flux}) is seen to correspond to $F_2$.
The question posed by the authors of \cite{IMU2} was how to explain
all the other moments. They suggested that this can be done in terms
of higher tensorial currents. In other words the Hawking radiation 
flows to infinity carried by higher tensor generalizations of the
energy--momentum tensor, which are coupled to suitable background 
fields that asymptotically vanish and do not back react.
 
The authors of \cite{IMU2,IMU4} used mostly higher spin currents
bilinear in a fermionic field. They also suggested an analogous 
construction with other kinds of fields, and briefly discussed the case
of a scalar bosonic field. In the following we would like to carry out the 
construction of higher spin currents in terms of a single complex
bosonic field ($c=2$). More explicitly, we will make use of the 
$W_\infty$ algebra constructed by Bakas and Kiritsis long ago,
\cite{BK}. To this end we go to the Euclidean and replace $u,v$ with
the complex coordinates $z, \bar z$.

\subsection{The $W_\infty$ algebra}

Following \cite{BK} (see also \cite{Bilal,Pope1,Pope2})
we start with free complex boson having the following two point functions
\begin{eqnarray} \label{eqtwop}
&&	\VEV{\phi(z_1)\overline{\phi}(z_2)} = 
-\log(z_1 - z_2)\CR
&&\VEV{\phi(z_1)\phi(z_2)} = 0\CR
&&	\VEV{\overline{\phi}(z_1)\overline{\phi}(z_2)} = 0
\end{eqnarray}

The currents are defined by
\begin{equation} \label{eqdefW}
	j^{(s)}_{z\ldots z}(z) = B(s) \sum_{k=1}^{s-1} (-1)^k A^s_k 
	\NO{\partial^k_z \phi(z) \partial^{s-k}_z \overline{\phi}(z)}
\end{equation}
where
\begin{equation} \label{eqdefB}
	B(s) = q^{s-2}\frac{2^{s-3}s!}{(2s-3)!!}
\end{equation}
and
\begin{equation} \label{eqdefA}
	A^s_k = \frac{1}{s-1}\binom{s-1}{k}\binom{s-1}{s-k}
\end{equation}
They satisfy a $W_\infty$ algebra. 
The first several currents are
\begin{eqnarray} \label{eqdefWex}
j^{(2)}_{zz} &=& -\NO{\partial_z \phi \partial_z \overline{\phi}} \CR
j^{(3)}_{zzz} &=&  -2 q \left( 
\NO{\partial_z \phi \partial^2_z \overline{\phi}}- 
\NO{\partial^2_z \phi \partial_z \overline{\phi}}\right) \CR
j^{(4)}_{zzzz} &=&  -\frac{16 q^2}{5} \left(
\NO{\partial_z \phi \partial^3_z \overline{\phi}} - 3\! 
\NO{\partial^2_z \phi \partial^2_z \overline{\phi}} + 
\NO{\partial^3_z \phi \partial_z \overline{\phi}} \right) \CR
j^{(5)}_{zzzzz} &=&  -\frac{32 q^3}{7} \left(
\NO{\partial_z \phi \partial^4_z \overline{\phi}} - 6\!
\NO{\partial^2_z \phi \partial^3_z \overline{\phi}} + 6\!
\NO{\partial^3_z \phi \partial^2_z \overline{\phi}} - 
\NO{\partial^4_z \phi \partial_z \overline{\phi}} \right) \CR
j^{(6)}_{zzzzzz} &=&  -\frac{128 q^4}{21} \left(
\NO{\partial_z \phi \partial^5_z \overline{\phi}} - 10\!
\NO{\partial^2_z \phi \partial^4_z \overline{\phi}} + 20\!
\NO{\partial^3_z \phi \partial^3_z \overline{\phi}} - 10\!
\NO{\partial^4_z \phi \partial^2_z \overline{\phi}} + 
\NO{\partial^5_z \phi \partial_z \overline{\phi}} \right)
\end{eqnarray}
Normal ordering is defined as
\begin{equation} \label{eqdefNO}
\NO{\partial^n \phi \partial^m \overline{\phi}} = 
\lim_{z_2 \rightarrow z_1} \left\{ 
\partial_{z_1}^n \phi(z_1) \partial_{z_2}^m \overline{\phi}(z_2) - 
\partial_{z_1}^n \partial_{z_2}^m \VEV{\phi(z_1)\overline{\phi}(z_2)}
\right\}
\end{equation}
As usual in the framework of conformal field theory, the operator 
product in the RHS is understood to be radial ordered.

The current $j^{(2)}_{zz}(z) = -\NO{\partial_z \phi(z) \partial_z \overline{\phi}(z)}$ is proportional to the (normalized) holomorphic 
energy-momentum tensor of the model
and, upon change of coordinates $z \rightarrow w(z)$, transforms as
\begin{equation} \label{eqW2trans}
\NO{\partial_z \phi \partial_z \overline{\phi}} = 
(w')^2\NO{\partial_w \phi \partial_w \overline{\phi}}
- \frac{1}{6 } \left\{ w, z \right\} 
\end{equation}
where $ \left\{ w, z \right\} $ --- the Schwarzian derivative --- is
\begin{equation} \label{eqSchwdef}
	\left\{ w, z \right\} = \frac{w'''(z)}{w'(z)} -
 \frac{3}{2}\left( \frac{w''(z)}{w'(z)} \right)^2 
\end{equation}
The non covariant contribution comes from the second term in \refb{eqdefNO}. 
We have (see e.g.\ \cite{Fabbri:2005mw})
\begin{eqnarray} \label{eq1}
\NO{\partial_{z_1} \phi(z_1) \partial_{z_2} \overline{\phi}(z_2)} &=&
\partial_{z_1} \phi(z_1) \partial_{z_2} \overline{\phi}(z_2)-
\partial_{z_1} \partial_{z_2} \VEV{\phi(z_1)\overline{\phi}(z_2)} \CR
&=&
w'(z_1)w'(z_2)\partial_{w_1} \phi(w_1) \partial_{w_2} \overline{\phi}(w_2)-
\partial_{z_1} \partial_{z_2} \VEV{\phi(z_1)\overline{\phi}(z_2)} \CR
&=&
w'(z_1)w'(z_2)\NO{\partial_{w_1} \phi(w_1) \partial_{w_2} \overline{\phi}(w_2)} 
-G(z_1,z_2)
\end{eqnarray}
where $\partial_{z_1} \phi(z_1) \partial_{z_2} \overline{\phi}(z_2)$ stands
for the radial ordered product of the two operators, and 
\begin{eqnarray} \label{eqdefG}\nonumber
G(z_1,z_2)
&=& 	
-w'(z_1)w'(z_2)\partial_{w_1} \partial_{w_2} \VEV{\phi(w_1)\overline{\phi}(w_2)} +
\partial_{z_1} \partial_{z_2} \VEV{\phi(z_1)\overline{\phi}(z_2)}
\CR
&=& 	
-\partial_{z_1} \partial_{z_2} \left(\VEV{\phi(w(z_1))\overline{\phi}(w(z_2))} -
\VEV{\phi(z_1)\overline{\phi}(z_2)}   \right)
\\
&=& 	
 \frac{w'(z_1)w'(z_2)}{(w(z_1)-w(z_2))^2}-
 \frac{1}{(z_1-z_2)^2}
\end{eqnarray}
In the limit $z_2 \rightarrow z_1$ \refb{eqdefG} becomes 
$\frac{1}{6 } \left\{ w, z_1 \right\}$.

We are interested in the transformation properties of currents $j^{(s)}(u)$ 
when $w(z)$ is 
\begin{equation}
	\label{eqdefw}
	w(z) = -e^{-\kappa z}
\end{equation}
Analogously to \refb{eq1}, we have
\begin{equation}\label{eq2}
j^{(s)}_{z\ldots z}(z_1) = 
\left(
B(s) \sum_{k=1}^{s-1} (-1)^k A^s_k 
\NO{\partial^k_{z_1} \phi(w(z_1)) \partial^{s-k}_{z_2} \overline{\phi}(w(z_2))}
\right)
+\VEV{ X}_s
\end{equation}
where
\begin{eqnarray} \label{defX2}
\VEV{X_s}
&=&  B(s) \sum_{k=1}^{s-1} (-1)^k A^s_k 
\lim_{z_2 \rightarrow z_1} \left\{ 
\VEV{\partial^k_{z_1} \phi(w(z_1)) \partial^{s-k}_{z_2} \overline{\phi}(w(z_2))} -
\VEV{\partial^k_{z_1} \phi(z_1) \partial^{s-k}_{z_2} \overline{\phi}(z_2)} 
\right\}
\CR
&=&  \lim_{z_2 \rightarrow z_1}
B(s) \sum_{k=1}^{s-1} (-1)^k A^s_k 
\partial^k_{z_1} \partial^{s-k}_{z_2} \left\{ 
\VEV{\phi(w(z_1)) \overline{\phi}(w(z_2))} -
\VEV{\phi(z_1) \overline{\phi}(z_2)} 
\right\}
\CR
&=&  \lim_{z_2 \rightarrow z_1}
B(s) \sum_{k=0}^{s-2} (-1)^{k+1} A^s_{k+1} 
\partial^k_{z_1} \partial^{s-k-2}_{z_2} \partial_{z_1} \partial_{z_2} \left\{ 
\VEV{\phi(w(z_1)) \overline{\phi}(w(z_2))} -
\VEV{\phi(z_1) \overline{\phi}(z_2)} 
\right\}
\CR
&=&  
B(s) \sum_{k=0}^{s-2} (-1)^{k} A^s_{k+1} 
\lim_{z_2 \rightarrow z_1} \partial^k_{z_1} \partial^{s-k-2}_{z_2} G(z_1,z_2)
\CR
&=&  
B(s) \sum_{k=0}^{s-2} (-1)^{k} A^s_{k+1} 
G_{k,s-k-2}
\end{eqnarray}
and $G_{m,n}$ are coefficients in the series
\begin{equation} \label{eqdefGmn}
	G(z+a,z+b)=\sum_{m,n=0}^{\infty} \frac{a^m b^n}{m! n!} G_{m,n}
\end{equation}
We now evaluate coefficients for the transformation \refb{eqdefw}. 
Putting \refb{eqdefw} in \refb{eqdefG} we obtain
\begin{equation} \label{eqGw}
G(z_1,z_2)=G(z_1-z_2)=
-\frac{1}{(z_1-z_2)^2} + 
\frac{\kappa^2}{4}\frac{1}{\sinh^2\frac{\kappa (z_1-z_2)}{2}}
\end{equation}
This gives\footnote{
Note that
\begin{eqnarray}\nonumber
-\frac{1}{ x^2} + 
\frac{\kappa^2}{4}\frac{1}{\sinh^2\frac{\kappa x}{2}}
&=&  \frac{d}{dx} \left( 
\frac{1}{x} \left( 1 - \frac{\kappa x}{e^{\kappa x}  - 1 }\right)
\right)
=  \frac{d}{dx} \left(  
\frac{1}{x} \left( 1 - \sum_{n=0}^{\infty} B_n \frac{(\kappa x)^n}{n!} \right)
\right)
\CR
&=& - \kappa^2
\sum_{n=2}^{\infty} (n-1) B_n \frac{(\kappa x)^{n-2}}{n!}
= - \kappa^2
\sum_{n=0}^{\infty} \frac{B_{n+2}}{n+2} \frac{(\kappa x)^{n}}{n!}
\end{eqnarray}
}
\begin{equation}
	G_{m,n}=(-)^{n+1}{\kappa^{m+n+2}}
\frac{B_{m+n+2}}{m+n+2}
\end{equation}
So, we obtain
\begin{eqnarray}\label{eqXw}
	 	\VEV{X_s}= (-)^{s-1} (4q)^{s-2}
	{\kappa^s} \frac{B_s}{s}
\end{eqnarray}
We have used 	
\begin{equation}
	\sum_{k=0}^{s-2} A^s_{k+1} = 
	\frac{(2s-2)!}{(s-1)!s!}	
\end{equation}

\refb{eqXw} is a higher order Schwarzian derivative evaluated at 
$w(z) = -e^{-\kappa z}$. It plays a role analogous to the RHS of 
\refb{flux}. In the next section we will compare it with the
radiation moments in the RHS of \refb{moments}.

\subsection{Higher moments of the black hole radiation}

Let us now return to the light--cone notation. We identify 
$j^{(2)}_{uu}(u)$ up to a 
constant\footnote{We relate $j^{(2)}_{uu}$ with the energy momentum 
tensor via the factor of $2\pi$ and the minus sign. This is because  
in the Euclidean we want to conform to the conventions and results of
\cite{BK}, where properly normalized currents satisfy a $W_\infty$ algebra.
This holds for higher order currents too: 
for physical applications their $W_\infty$ representatives must 
all be divided by $-2\pi$.} with
the holomorphic energy momentum tensor 
\be
 j_{uu}^{(2)}(u)=-2\pi\, T^{(hol)}_{uu}\label{j2T}
\ee
Similarly we identify 
$j^{(s)}_{u\ldots u}$, with $s$ lower 
indices, with an $s$--th order holomorphic tensor. They can be 
naturally thought of as the only
non--vanishing components of a two--dimensional completely 
symmetric current. In analogy with the
energy--momentum tensor, we expect that there exist
a conformally covariant version $J^{(s)}_{u\ldots u}$ of 
$j^{(s)}_{u\ldots u}$. The latter must be the intrinsic 
component of a two--dimensional completely symmetric traceless current
$ J^{(s)}_{\mu_1\ldots\mu_s}$, whose only other classically non--vanishing
component is $J^{(s)}_{v\ldots v}$. 

Now let us apply to these currents an argument similar to the 
one in section 2 
for the energy--momentum tensor, using the previous results from 
the $W_\infty$ algebra. Introducing the Kruskal coordinate $U=-e^{-\kappa u}$ and 
requiring regularity at the horizon we find that, at the horizon, the value
of $j^{(s)}_{u\ldots u}$ is given by $\VEV{X_s}$ in eq.\refb{eqXw}. Next
$j^{(s)}_{u\ldots u}(u)$ is constant in $t$ and $r$ (the same is of course true
for $j^{(s)}_{v\ldots v}$). Therefore, if we identify 
$j^{(s)}_{u\ldots u}(u)$ with $j^{(s)}_{z\ldots z}(z)$ via Wick rotation,
$\VEV{X_s}$ corresponds to its value at $r=\infty$. Since $j^{(s)}_{u\ldots u}(u)$
and  $J^{(s)}_{u\ldots u}(u)$ asymptotically coincide, the asymptotic
flux of this current is 
\be
-\frac 1{2\pi} \VEV{J^{(s)^r}{}_{t\ldots t}}= -\frac 1{2\pi} \VEV{J^{(s)}_{u\ldots u}}+
\frac 1{2\pi} \VEV{J^{(s)}_{v\ldots v}}= -\frac 1{2\pi} \VEV{X_s}= 
\frac {i^{s-2}}{2\pi s} \kappa^s B_s \label{sflux}
\ee
provided we set the deformation parameter $q$ to the value $-\frac i4$
(for the global $-2\pi$ factor, see the previous footnote). 

The RHS vanishes
for odd $s$ (except $s=1$ which is not excited in our case) and coincides 
with the thermal flux moments \refb{moments} for even $s$.
 
It remains for us to show that the covariant conserved currents  
$ J^{(s)}_{\mu_1\ldots\mu_s}$ can be defined.

\section{Higher spin covariant currents}

To start with, it is natural to suppose that the covariant currents
appear in an effective action $S$ where they are    
sourced by asymptotically trivial
background fields $B^{(s)}_{\mu_1\ldots\mu_s}$ (in \cite{hull} they were 
called `cometric functions'), i.e.
\be
 J^{(s)}_{\mu_1\ldots\mu_s}= \frac 1{\sqrt{g}} \frac {\delta}{\delta 
B^{(s)\mu_1\ldots\mu_s}} S\label{Js}
\ee
In particular $B^{(2)}_{\mu\nu}= g_{\mu\nu}/2$. We assume that
all $J^{(s)}_{\mu_1\ldots\mu_s}$ are maximally symmetric and classically
traceless.

In order to find a covariant expression we first recall that the previous
$W_\infty$ algebra is formulated in terms of a (complex, Euclidean) 
chiral bosonic field. The action of a chiral (Minkowski) scalar in 2D 
coupled to background 
gravity can be found in \cite{Sonn88}. When the background gravity
is of the type considered in this paper, i.e. 
$g_{\alpha\beta} = e^\varphi \eta_{\alpha\beta}$, 
the action boils down to 
that of a free chiral boson, \cite{Flo87}. In other words, the equation
of motion of a chiral boson coupled to background conformal gravity is
\be
\partial_v \phi =0\label{EoMphi}
\ee
This simplifies the covariantization process. 

To proceed with the covariantization program we then
reduce the problem to a one--dimensional one. We consider only the
$u$ dependence and keep $v$ fixed. In one dimension a curved 
coordinate $u$ is easily related to the corresponding normal 
coordinate $x$ via the relation $\partial_x = e^{-\varphi(u)} \partial_u$.
 We view
$u$ as $u(x)$, assume that all $j^{(s)}_{u\ldots u}$ and their $W_\infty$
relations refer in fact to the flat $x$ 
coordinate (i.e. $x$ corresponds to the Euclidean coordinate $z$ used
in the previous section)
and by the above equivalence we extract the components in the
new coordinate system. For instance for a scalar field $\phi$:
\be
\partial_x^n \phi = e^{-n\varphi(u)} \nabla^n_u \phi,\quad\quad{\rm i.e.}
\quad\quad
\partial_x^n \phi\, (dx)^n = \nabla^n_u\phi\, (du)^n\0
\ee
We recall that the $W_\infty$ currents are constructed out of
bilinears in $\phi$ and $\bar \phi$: 
\begin{equation}
	j^{(n,m)}_{u\ldots u} = \NO{\partial^n_u \phi \partial^m_u \overline{\phi}}\label{jnm}
\end{equation}
We split the factors and evaluate one factor in $u_1 = u(x+\epsilon/2)$ 
and the other in
$u_2 = u(x-\epsilon/2)$. We expand in $\epsilon$ and take the 
limit for $\epsilon \to 0$. Afterward we restore the tensorial
character of the product by multiplying it by a suitable 
$e^{n\varphi(u)}$ factor. We use in particular the Taylor expansion, 
see \cite{IMU4},
\be
u(x+\epsilon)= u(x)+ \epsilon \,e^{-\varphi}- \frac {\epsilon^2} 2
e^{-2\varphi}\,\partial_u \varphi+\ldots\0
\ee

According to the recipe just explained, the covariant counterpart
of  $j^{(s)}_{u\ldots u}$ should be constructed using currents
\begin{equation}
J^{(n,m)}_{u\ldots u} = e^{(n+m)\varphi(u)}\lim_{\epsilon \rightarrow 0} 
\left\{ e^{-n\varphi(u_1)-m\varphi(u_2)} 
\nabla^n_{u_1} \phi \nabla^m_{u_2} \overline{\phi} - 
\frac{c_{n,m} \hbar}   {\epsilon^{n+m}}\right\}\label{Jnm}
\end{equation}
where $c_{n,m} = (-)^{m} ( n+m-1 )!$ are numerical constants determined in such a way
that all singularities are canceled in the final expression for $J^{(n,m)}_{u\ldots u}$.
Therefore
\refb{Jnm} defines the normal ordered current
\begin{equation}
	J^{(n,m)}_{u\ldots u} = \NO{\nabla^n_u \phi \nabla^m_u 
\overline{\phi}}\label{Jnmno}
\end{equation}

We use
\be
\nabla_u \nabla_u^n f(u,v) = \partial_u \nabla_u^n f(u,v) - 
n \Gamma \nabla_u^n f(u,v) \nonumber
\ee
for a scalar field $f(u,v)$, where
\begin{equation}
	\Gamma = \partial_u \varphi \nonumber
\end{equation}
and
\begin{equation}
\NO{\phi(u_1)\overline{\phi}(u_2)} = \phi(u_1)\overline{\phi}(u_2) + 
\hbar\log(u_1 - u_2)
\end{equation}
After some algebra we obtain 
\begin{eqnarray}\label{Jn,m}
 J^{(1,1)}_{uu}&=&\frac{ \hbar }{6}T+j^{(1,1)}_{uu} \CR
 J^{(1,2)}_{uuu}&=&\frac{\hbar  }{12}\left(\partial _uT\right)-
\Gamma  J^{(1,1)}_{uu}+j^{(1,2)}_{uuu} \CR
 J^{(2,1)}_{uuu}&=&\frac{\hbar  }{12}\left(\partial _uT\right)-
\Gamma  J^{(1,1)}_{uu}+j^{(2,1)}_{uuu} \CR
 J^{(1,3)}_{uuuu}&=&\frac{\hbar  }{20 }\left(\partial _u^2T\right)
+\frac{\hbar }{30} T^2-J^{(1,1)}_{uu} T-\frac{3}{2} \Gamma ^2 J^{(1,1)}_{uu}-3 
\Gamma  J^{(1,2)}_{uuu}+j^{(1,3)}_{uuuu} \CR
 J^{(2,2)}_{uuuu}&=&\frac{\hbar  }{30 } \left(\partial _u^2T\right)
-\frac{\hbar  }{30 }T^2-\Gamma ^2 J^{(1,1)}_{uu}-\Gamma  J^{(1,2)}_{uuu}
-\Gamma  J^{(2,1)}_{uuu}+j^{(2,2)}_{uuuu} \CR
 J^{(3,1)}_{uuuu}&=&\frac{\hbar  }{20 }\left(\partial _u^2T\right)
+\frac{\hbar  }{30 }T^2-J^{(1,1)}_{uu} T-\frac{3}{2} \Gamma ^2 J^{(1,1)}_{uu}
-3 \Gamma  J^{(2,1)}_{uuu}+j^{(3,1)}_{uuuu}
\end{eqnarray}
where
\begin{equation}
T = \partial^2_u \varphi -  \frac{1}{2} 
\left( \partial_u \varphi \right)^2  
\end{equation}
In Appendix one can find analogous expressions for order 5 and 6 currents.

Using Eq.~\refb{eqdefWex}, and similarly,
$J^{(2)}_{uu} = -J^{(1,1)}_{uu}$,\,
$J^{(3)}_{uuu} = -2 q \left( J^{(1,2)}_{uuu} - J^{(2,1)}_{uuu} \right)$,\,
$J^{(4)}_{uuuu} = -\frac{16 q^2}{5} \left( J^{(1,3)}_{uuuu} - 3 J^{(2,2)}_{uuuu} 
+ J^{(3,1)}_{uuuu} \right)$, etc.,
we obtain
\begin{eqnarray}\label{Jsjs}
	J^{(2)}_{uu} &=& j^{(2)}_{uu}  - \frac{ \hbar }{6}  T  \CR
	J^{(3)}_{uuu} &=&  j^{(3)}_{uuu}  \CR
	J^{(4)}_{uuuu} &=&  j^{(4)}_{uuuu}  - \frac{ 8\hbar}{15}  q^2 T^2 - 
\frac{32}{5}q^2TJ^{(2)}_{uu}   \CR 
	J^{(5)}_{uuuuu} &=&  j^{(5)}_{uuuuu} - \frac{160}{7}  q^2 T  J^{(3)}_{uuu}  
\end{eqnarray}
For $s=6$ we have
\begin{eqnarray}\label{J6j6}
	J^{(6)}_{uuuuuu} &=&  \left(
-\frac{512 \hbar }{63  }T^3
+\frac{160 \hbar }{63  }\left(\partial _uT\right)^2
-\frac{128 \hbar }{63  }T\partial _u^2T\right. 
\CR
&&
-\frac{512}{3}T^2J^{(2)}_{uu}
-\frac{256}{21}T\nabla _u^2J^{(2)}_{uu}
-\frac{256}{21}\left(\partial _u^2T\right)J^{(2)}_{uu}
+\frac{640}{21}\left(\partial _uT\right)\nabla _uJ^{(2)}_{uu}
\CR
&&
\left.
-\frac{1280}{21}\Gamma T\nabla _uJ^{(2)}_{uu}
-\frac{1280}{21}\Gamma ^2TJ^{(2)}_{uu}
+\frac{1280}{21}\Gamma \left(\partial _uT\right)J^{(2)}_{uu}\right) q^4
-\frac{160}{3}q^2TJ^{(4)}_{uuuu}
+j^{(6)}_{uuuuuu}
\end{eqnarray}
 
It is important to verify that our previous definitions
are consistent. 
Using the transformation law for $j^{(2)}_{uu}$ (i.e.\ \refb{eqW2trans})
\begin{equation}
	j^{(2)}_{uu}(u) = (w'(u))^2 \,\tilde{j}^{(2)}_{ww}(w(u)) + 
\frac{\hbar}{6} \left\{ w, u \right\} 
\end{equation}
and its generalization for  $j^{(4)}_{uuuu}$ which can be read out of \refb{eq2}
\begin{equation}
	j^{(4)}_{uuuu}(u) = (w'(u))^4 \,\tilde{j}^{(4)}_{wwww}(w(u)) + 
	q^2 (w')^2\, \frac{32}{5}\, \tilde{j}^{(2)}_{ww}(w(u))
 \left\{ w, u \right\} +
	\hbar q^2\,  \frac{8}{15 } \, \left\{ w, u \right\}^2
\end{equation}
and using 
\begin{equation}
	\varphi(u,v) = \tilde{\varphi}(w(u),v)+ \log( w'(u) )
\end{equation}
it can be checked that $J^{(2)}_{uu}$ and $J^{(4)}_{uuuu}$ transform indeed as tensors
\begin{eqnarray}
	J^{(2)}_{uu}(u) &=& (w'(u))^2 \tilde{J}^{(2)}_{ww} ( w(u)) \CR
	J^{(4)}_{uuuu}(u) &=&  (w'(u))^4	  \tilde{J}^{(4)}_{wwww} (w(u))
\end{eqnarray}

The next step consists in finding the covariant derivatives of the 
currents. The only $v$ dependence comes from $\varphi$. We have
\begin{eqnarray}
g^{uv}\nabla _vJ^{(1,1)}_{uu}&=&-\frac{\hbar  }{12}
\left(\nabla _uR\right) \CR
 g^{uv}\nabla _vJ^{(1,2)}_{uuu}&=&-\frac{\hbar  }{24}
\left(\nabla _u^2R\right) + \frac{1}{2} R J^{(1,1)}_{uu} \CR
 g^{uv}\nabla _vJ^{(2,1)}_{uuu}&=&-\frac{\hbar  }{24 }
\left(\nabla _u^2R\right) + \frac{1}{2} R J^{(1,1)}_{uu} \CR
 g^{uv}\nabla _vJ^{(1,3)}_{uuuu}&=&-\frac{\hbar  }{40 }
\left(\nabla _u^3R\right)+\frac{1}{2} \left(\nabla _uR\right) J^{(1,1)}_{uu}+
\frac{3}{2} R J^{(1,2)}_{uuu} \CR
 g^{uv}\nabla _vJ^{(2,2)}_{uuuu}&=&-\frac{\hbar  }{60 }
\left(\nabla _u^3R\right)+\frac{1}{2} R J^{(1,2)}_{uuu}+\frac{1}{2} R J^{(2,1)}_{uuu} \CR
 g^{uv}\nabla _vJ^{(3,1)}_{uuuu}&=&-\frac{\hbar  }{40 }
\left(\nabla _u^3R\right)+\frac{1}{2} \left(\nabla _uR\right) J^{(1,1)}_{uu}
+\frac{3}{2} R J^{(2,1)}_{uuu}
\end{eqnarray}
and, using \refb{J5} and \refb{J6} in Appendix,
\begin{eqnarray}\nonumber
 g^{uv}\nabla _vJ^{(1,4)}_{uuuuu} &=& 
-\frac{\hbar  }{60 }\left(\nabla _u^4R\right)
+\frac{1}{2} \left(\nabla _u^2R\right) J^{(1,1)}_{uu}
+2 \left(\nabla _uR\right) J^{(1,2)}_{uuu}+3 R J^{(1,3)}_{uuuu} \CR
 g^{uv}\nabla _vJ^{(2,3)}_{uuuuu} &=& -\frac{\hbar  }{120 }
\left(\nabla _u^4R\right)+\frac{1}{2} \left(\nabla _uR\right) J^{(2,1)}_{uuu}
+\frac{1}{2} R J^{(1,3)}_{uuuu}+\frac{3}{2} R J^{(2,2)}_{uuuu} \CR
 g^{uv}\nabla _vJ^{(3,2)}_{uuuuu} &=& -\frac{\hbar  }{120 }
\left(\nabla _u^4R\right)+\frac{1}{2} \left(\nabla _uR\right) J^{(1,2)}_{uuu}
+\frac{1}{2} R J^{(3,1)}_{uuuu}+\frac{3}{2} R J^{(2,2)}_{uuuu} \CR
 g^{uv}\nabla _vJ^{(4,1)}_{uuuuu} &=& -\frac{\hbar  }{60 }
\left(\nabla _u^4R\right)+\frac{1}{2} \left(\nabla _u^2R\right) J^{(1,1)}_{uu}
+2 \left(\nabla _uR\right) J^{(2,1)}_{uuu}+3 R J^{(3,1)}_{uuuu}
\end{eqnarray}
\begin{eqnarray}\nonumber
 g^{uv}\nabla _vJ^{(1,5)}_{uuuuuu} &=& -\frac{\hbar  }{84 }
\left(\nabla _u^5R\right)+\frac{1}{2} \left(\nabla _u^3R\right) J^{(1,1)}_{uu}
+\frac{5}{2} \left(\nabla _u^2R\right) J^{(1,2)}_{uuu}
+5 \left(\nabla _uR\right) J^{(1,3)}_{uuuu}+5 R J^{(1,4)}_{uuuuu} \CR
 g^{uv}\nabla _vJ^{(2,4)}_{uuuuuu} &=& -\frac{\hbar  }{210 }
\left(\nabla _u^5R\right)+\frac{1}{2} \left(\nabla _u^2R\right) J^{(2,1)}_{uuu}
+2 \left(\nabla _uR\right) J^{(2,2)}_{uuuu}+\frac{1}{2} R J^{(1,4)}_{uuuuu}+3 R J^{(2,3)}_{uuuuu} \CR
 g^{uv}\nabla _vJ^{(3,3)}_{uuuuuu} &=& -\frac{\hbar  }{280 }
\left(\nabla _u^5R\right)+\frac{1}{2} \left(\nabla _uR\right) J^{(1,3)}_{uuuu}
+\frac{1}{2} \left(\nabla _uR\right) J^{(3,1)}_{uuuu}+\frac{3}{2} R J^{(2,3)}_{uuuuu}
+\frac{3}{2} R J^{(3,2)}_{uuuuu} \CR
 g^{uv}\nabla _vJ^{(4,2)}_{uuuuuu} &=& -\frac{\hbar  }{210 }
\left(\nabla _u^5R\right)+\frac{1}{2} \left(\nabla _u^2R\right) J^{(1,2)}_{uuu}
+2 \left(\nabla _uR\right) J^{(2,2)}_{uuuu}+\frac{1}{2} R J^{(4,1)}_{uuuuu}+3 R J^{(3,2)}_{uuuuu} \CR
 g^{uv}\nabla _vJ^{(5,1)}_{uuuuuu} &=& -\frac{\hbar  }{84 }
\left(\nabla _u^5R\right)+\frac{1}{2} \left(\nabla _u^3R\right) J^{(1,1)}_{uu}
+\frac{5}{2} \left(\nabla _u^2R\right) J^{(2,1)}_{uuu}+
5 \left(\nabla _uR\right) J^{(3,1)}_{uuuu}+5 R J^{(4,1)}_{uuuuu}
\end{eqnarray}

For the  currents $J^{(s)}_{u\ldots u}$, which are the linear combinations of $J^{(n,m)}_{u\ldots u}$ we obtain
\begin{eqnarray}
 g^{uv}\nabla _v	J^{(2)}_{uu} &=&  
\frac{ \hbar }{12} \left(\nabla _uR\right) 
\label{djv2}\\
g^{uv}\nabla _v J^{(3)}_{uuu} &=& 0
\label{djv3}\\
 g^{uv}\nabla _v	J^{(4)}_{uuuu} &=&  \frac{ 16}{5}  q^2 
\left(\nabla _uR\right) J^{(2)}_{uu}  
\\
	g^{uv}\nabla _v  J^{(5)}_{uuuuu}  &=&  \frac{80}{7} q^2 \left(\nabla _uR\right) J^{(3)}_{uuu}
\end{eqnarray}
For  $s=6$:
\begin{eqnarray}\nonumber
g^{uv}\nabla _v  J^{(6)}_{uuuuuu}  &=& 
\left(-\frac{320}{21}\left(\nabla _u^2R\right)\nabla _uJ^{(2)}_{uu}
+\frac{128}{21}\left(\nabla _uR\right)\nabla _u^2J^{(2)}_{uu}
+\frac{128}{21}\left(\nabla _u^3R\right)J^{(2)}_{uu}\right) q^4 
\label{djv6}\\
&+& \frac{80}{3}\left(\nabla _uR\right)J^{(4)}_{uuuu} q^2
\end{eqnarray}

Now, according to \cite{IMU4},
after the right hand side is expressed in terms of covariant quantities, 
terms proportional to $\hbar$ are identified as anomalies in the following 
way. One assumes that there is no anomaly in the conservation laws of covariant 
currents, i.e\  that the terms proportional to $\hbar$ do not appear in 
$\nabla^\mu  J_{\mu u\ldots u}$. Since $\nabla^\mu  J_{\mu u\ldots u} = g^{uv} 
\nabla_v J_{uu\ldots u} + g^{uv} \nabla_u J_{vu\ldots u}$, one relates 
terms proportional to $\hbar$  in the $u$ derivative of the trace 
($_{vu\ldots u}$ components) with the terms proportional to $\hbar$ in 
the $v$ derivative of $_{uu\ldots  u}$ components of the currents.

For the covariant energy momentum tensor $J^{(2)}_{\mu\nu}$, the trace is  
$\Tr( J^{(2)}{} ) = 2 g^{vu}J^{(2)}{}_{vu} $. Thus, \refb{djv2} reproduces the well 
known trace anomaly $ \Tr( J^{(2)} ) = -\frac{c \hbar}{12} R$, where 
in our case $c=2$ (for the missing factor of $-2\pi$ see the footnote
in section 2.2).

We see that the terms that carry explicit factors of $\hbar$ cancel out
in eqs.~\refb{djv3}-\refb{djv6}. This implies the absence of $\hbar$ 
in the trace, and consequently the absence of the trace anomaly.

\section{Trace anomalies}

In the previous section the covariant form of the current does not 
give rise to any trace anomaly. This is at variance with ref.\cite{IMU4},
where the fourth order covariantized current exhibits a trace 
anomaly which is a superposition of three terms: 
$\nabla_{\mu}\nabla_{\nu}R,\,
g_{\mu\nu}\square R$ and $g_{\mu\nu}R^2$. It is therefore important to clarify whether
these are true anomalies or whether they are some kind of artifact of the 
regularization used to derive the results. 

In the framework of the effective action introduced in the previous 
section (see \refb{Js}), the anomaly problem can be clarified using 
cohomological (or consistency) methods. 
Such methods were applied for the first time to the study of trace 
anomalies in \cite{Bonora83,Bonora85}. Subsequent applications can be 
found in \cite{Bonora84,Deser} and more recently in 
\cite{Boulanger1,Boulanger2}. 
The consistency conditions for trace anomalies are similar to the 
Wess--Zumino consistency conditions 
for chiral anomalies and are based on the simple remark that, if we 
perform two symmetry transformations in different order on the one--loop 
action, the result must obey the group theoretical rules of the 
transformations. In particular, since Weyl
transformations are Abelian, making two Weyl transformations in opposite 
order must bring the same result. Although this explains the geometrical 
meaning of the consistency conditions, proceeding in this way is often very 
cumbersome.
The problem becomes more manageable if we transform it into a cohomological 
one. This is simple: just promote the local transformation parameters to 
anticommuting fields (ghost). The transformations become nilpotent and 
define a coboundary operator.

In this section we will consider, for simplicity, the possible anomalies 
of the fourth order current $J^{(4)}_{\mu\nu\la\rho}$ which couples in 
the action to the background field 
$B^{(4)}_{\mu\nu\la\rho}\equiv B_{\mu\nu\la\rho} $,
both being completely symmetric tensors. The relevant Weyl 
transformations are as follows. The gauge parameters are the usual Weyl 
parameter $\sigma$ and 
new Weyl parameters $\tau_{\mu\nu}$ (symmetric in $\mu,\nu$). 
The variation $\delta_\tau$ acts only on $B_{\mu \nu\lambda\rho}$
(see \cite{hull})
\begin{equation}
\delta_\tau B_{\mu \nu \lambda\rho} = 
g_{\mu \nu}\,\tau_{ \lambda\rho} +
g_{\mu \lambda}\,\tau_{\nu\rho} +
g_{\mu \rho}\,\tau_{\nu\lambda} +
g_{\nu\lambda }\,\tau_{\mu  \rho} +
g_{ \nu\rho}\,\tau_{\mu \lambda} +
g_{\lambda\rho }\,\tau_{\mu \nu} \label{tautransf}
\end{equation}
while $\delta_\sigma$ acts on $g_{\mu\nu}$, 
$\tau_{\mu\nu}$ and $B_{\mu \nu \lambda\rho}$ in the following way
\begin{eqnarray} 
&& \delta_\sigma g_{\mu \nu}  = 2 \,\sigma\, g_{\mu \nu}  \CR
&& \delta_\sigma \tau_{\mu \nu}  = (x-2)\,\sigma\, \tau_{\mu \nu} \CR
&& \delta_\sigma B_{\mu \nu\lambda \rho} = x\,\sigma\, 
B_{\mu \nu\lambda\rho}
\label{sigmatransf}
\end{eqnarray}
where $x$ is a free numerical parameter. The transformation 
\refb{sigmatransf}
of $\tau$ and $B$ are required for consistency with \refb{tautransf}. 
The actual value of $x$ turns out to be immaterial.

Now we promote $\sigma$ and $\tau$ to anticommuting fields:
\begin{eqnarray} 
&& \sigma^2 = 0\0 \\
&& \tau_{\mu \nu}\, \tau_{\lambda \rho}+\tau_{\lambda \rho}\,\tau_{\mu \nu} 
= 0\0 \\
&& \sigma\,\tau_{\mu \nu}  +\tau_{\mu \nu}\, \sigma=0\0
\end{eqnarray}
It is easy to verify that
\be
\delta_\sigma^2=0,\quad\quad \delta_\tau^2=0,\quad\quad \delta_\sigma\, 
\delta_\tau+ \delta_\tau\, \delta_\sigma=0\0
\ee
Therefore they define a double complex. 

Integrated anomalies are defined by 
\be
\delta_\sigma \Gamma^{(1)} = \hbar\, \Delta_\sigma,
\quad\quad \delta_\tau \Gamma^{(1)} = \hbar\, \Delta_\tau,\label{cocycles}
\ee
where $\Gamma^{(1)}$ is the one--loop quantum action and 
$\Delta_\sigma,\Delta_\tau$ are
local functional linear in $\sigma$ and $\tau$, respectively. The 
unintegrated anomalies, i.e. the traces $T_\mu^\mu$ and 
$J^{(4)\mu}{}_{\mu\la\rho}$ are obtained
by functionally differentiating with respect to 
$\sigma$ and $\tau_{\la\rho}$,
respectively.

By applying $\delta_\sigma,\delta_\tau$ to the eqs.\refb{cocycles}, we see 
that candidates for anomalies  $\Delta_\sigma$ and $\Delta_\tau$
must satisfy the consistency conditions
\begin{equation}\label{condss}
\delta_\sigma\, \Delta_\sigma  = 0
\end{equation}
\begin{equation}\label{condst}
\delta_\tau\, \Delta_\sigma + \delta_\sigma \,\Delta_\tau = 0
\end{equation}
\begin{equation}\label{condtt}
\delta_\tau\,\Delta_\tau = 0
\end{equation}
i.e. they must be cocycles. We have to make sure that they are true 
anomalies, that is that they are nontrivial. In other words there must 
not exist local counterterm $C$ in the action such that 
\begin{eqnarray} \label{condtriv1}
	\Delta_\sigma &=& \delta_\sigma\, \int d^2x   \,C\\ 
\label{condtriv2}
\Delta_\tau &=&  \delta_\tau\, \int d^2x\,   C
\end{eqnarray}
If such a $C$ existed we could redefine the quantum action by subtracting these 
counterterms and get rid of the (trivial) anomalies.

We start by expanding candidate anomalies as linear combinations of 
curvature invariants\footnote{
The fact that we are in $2$ spacetime dimensions reduces greatly the 
number of curvature invariants, such as those in \refb{defI}.
Useful relations valid in $2$ dimensions are
\begin{eqnarray}
	R_{\mu \nu \lambda \rho} &=&  \frac{1}{2} R 
\left( g_{\mu \lambda}\, g_{\nu \rho} 
- g_{\mu \rho}\, g_{\nu \lambda} \right)  \CR
	R_{\mu \nu}  &=&  \frac{1}{2} g_{\mu \nu}\, R \CR
\delta_\sigma R &=&  -2\, R\, \sigma - 2\, \square \sigma
\end{eqnarray}
}
\begin{equation}\label{defDs}
	\Delta_\sigma = \int d^2x\, \sqrt{-g} \sum_{i=2}^{11} c_i\, I_i
\end{equation}
\begin{equation}\label{defDt}
	\Delta_\tau =  \int d^2x, \sqrt{-g} \sum_{k=1}^{3} b_k\, K_k
\end{equation}
where $I_i$ are linear in 
$B^{\mu\nu\lambda\rho}$ 
and $\sigma$:
\begin{eqnarray}	\label{defI}
       I_1   &=& \sigma R\CR
	I_2    &=& B^{\mu\nu\la\rho}\,\,
\nabla_\mu\nabla_\nu\nabla_\lambda\nabla_\rho\sigma	\CR
	I_3    &=& B^{\mu \nu} \,R\,\,\nabla_\mu\nabla_\nu\sigma\CR
	I_4    &=& B^{\mu \nu}\,\, \nabla_\mu\nabla_\nu\square\sigma\CR
	I_5    &=& B^{\mu \nu} \,\, \nabla_\mu\nabla_\nu R \sigma \CR
	I_6    &=& B\, \square R\,\,\sigma	\CR
	I_7    &=& B\, R^2\, \sigma	\CR
	I_8    &=& B^{\mu \nu}\,   \nabla_\mu \,R\,  \nabla_\nu\sigma 	\CR
	I_9    &=& B\, R\, \square\sigma 	\CR
	I_{10} &=& B\, g^{\mu \nu} \,\, \nabla_\mu R \,\,\nabla_\nu\sigma\CR
	I_{11} &=& B \,\,\square^2 \sigma
\end{eqnarray}
($B^{\mu \nu}  = B^{\mu\nu\lambda\rho } g_{\lambda\rho}$, 
$B = B^{\mu\nu}  g_{\mu\nu} $). The term $I_1$ corresponds to the usual anomaly of 
the energy--momentum trace (which is consistent and nontrivial). Therefore in the 
sequel we disregard it and limit ourselves to the other terms which contain 
4 derivatives. Similarly $K_k$ are
independent curvature invariants that are linear in $\tau_{\mu \nu} $ and contain 
4 derivatives:
\begin{eqnarray}	\label{defK}
	K_1    &=& \nabla_\mu\nabla_\nu R\,\, \tau^{\mu \nu} \CR
	K_2    &=&  R^2\,	\tau\CR
	K_3    &=& \square R \,\,\tau 
\end{eqnarray}
where $\tau = g^{\mu \nu} \, \tau_{\mu \nu} $.

Now we apply the consistency condition \refb{condss} to  $\Delta_\sigma$
in the form \refb{defDs}. We obtain
\begin{equation}
	\delta_\sigma \Delta_\sigma =
	 \sum_{i=2}^{11} \sum_{j=1}^{12}  c_i A_{ij}\int d^2x\,\sqrt{-g} 
J^{\sigma\sigma}_j = 0
\end{equation}
where the variations $\delta_\sigma I_i$ are expressed as linear combinations of 
terms $J^{\sigma\sigma}_j$ 
\begin{eqnarray}	\label{defJ}
J^{\sigma\sigma}_{1} &=& B^{\mu \nu}\,R \,\sigma \,
\nabla_\mu\nabla_\nu\sigma\CR
J^{\sigma\sigma}_{2} &=& B^{\mu \nu}\,\nabla_\mu R \,
\sigma \,\nabla_\nu\sigma\CR
J^{\sigma\sigma}_{3} &=& B \,R \,\sigma\,\square \sigma	\CR
J^{\sigma\sigma}_{4} &=& B\, g^{\mu \nu}  \,\nabla_\mu R \,\sigma\,\, 
\nabla_\nu\sigma	\CR
J^{\sigma\sigma}_{5} &=& 
B^{\mu\nu\lambda\rho}\,\sigma\,
\nabla_\mu\nabla_\nu\nabla_\lambda\nabla_\rho\sigma\CR
J^{\sigma\sigma}_{6} &=& 
B^{\mu \nu\lambda\rho }\,\nabla_\mu\sigma 
\,\nabla_\nu\nabla_\lambda\nabla_\rho\sigma\CR
J^{\sigma\sigma}_{7} &=& 
B^{\mu \nu}\, \sigma\,\nabla_\mu\nabla_\nu\square\sigma	\CR
J^{\sigma\sigma}_{8} &=& 
B^{\mu \nu}\, \nabla_\mu\sigma \,\nabla_\nu\square\sigma\CR
J^{\sigma\sigma}_{9} &=& 
B^{\mu \nu}\, \square\sigma\, \nabla_\mu\nabla_\nu\sigma \CR
J^{\sigma\sigma}_{10} &=&
 B^{\mu \nu}\, g^{\lambda\rho }\, \nabla_\lambda\sigma 
\,\nabla_\mu\nabla_\nu\nabla_\rho\sigma \CR
J^{\sigma\sigma}_{11} &=& B\,\sigma\,\square^2\sigma\CR
J^{\sigma\sigma}_{12} &=& 
B\,g^{\mu \nu}\,\nabla_\mu\sigma\, \nabla_\nu\square\sigma
\end{eqnarray}
with coefficients given by
\begin{equation} \label{explA}
	A_{ij} = \left( \begin{array}{cccccccccccc}
 0 & 0 & 0 & 0 & x-6 & -10 & 0 & 0 & 0~~ & 5 & 0 & 0 \\
 x-6 & 0 & 0 & 0 & 0 & 0 & 0 & 0 & -2 & 0 & 0 & 0 \\
 0 & 0 & 0 & 0 & 0 & 0 & x-6 & -6 & 2 & 0 & 0 & 1 \\
 2 & 6 & 0 & -1 & 0 & 0 & 2 & 0 & 0 & 0 & 0 & 0 \\
 0 & 0 & 2 & 4 & 0 & 0 & 0 & 0 & 0 & 0 & 2 & 0 \\
 0 & 0 & 4 & 0 & 0 & 0 & 0 & 0 & 0 & 0 & 0 & 0 \\
 0 & x-6 & 0 & 0 & 0 & 0 & 0 & 2 & 0 & 0 & 0 & 0 \\
 0 & 0 & x-6 & 0 & 0 & 0 & 0 & 0 & 0 & 0 & 0 & 0 \\
 0 & 0 & 0 & x-6 & 0 & 0 & 0 & 0 & 0 & 0 & 0 & 2 \\
 0 & 0 & 0 & 0 & 0 & 0 & 0 & 0 & 0 & 0 & x-6 & -4
\end{array} \right)
\end{equation}
This gives a homogeneous system of equations for $c_2$, \ldots, $c_{11}$
\begin{equation}
	\sum_{i=2}^{11}  c_i A_{ij} = 0\;,\quad j=1,\ldots,12 
\end{equation}
The solution can be expressed in terms of 3 free parameters which we take to
 be $c_9$, $c_{10}$, $c_{11}$.
We have
\begin{eqnarray}\label{solc}
c_2&=&0\CR
c_3&=&-2 \left(c_{10}-2 c_{11}\right)\CR
c_4&=&-2 \left(c_{10}-2 c_{11}\right)\CR
c_5&=& (c_{10}- 2 c_{11} )(x-6)\CR
c_6&=&-\frac{1}{2} c_{11} (x-6)\CR
c_7&=&\frac{1}{4} (x-6) (c_{11} - c_9)  \CR
c_8&=&-6 \left(c_{10}-2 c_{11}\right)
\end{eqnarray}

Now we plug this solution \refb{solc} back into \refb{defDs} and apply the 
consistency condition \refb{condst}  
\begin{eqnarray}\nonumber
\delta_\tau \Delta_\sigma + \delta_\sigma \Delta_\tau &=& 
\delta_\tau \left( \int d^2x \sqrt{-g} \sum_{i=2}^{12}c_i I_i \right) + 
\delta_\sigma \left( \int d^2x \sqrt{-g} \sum_{k=1}^{3}b_k K_k \right) 
\\ \label{e18}
&=& 
\int d^2x \sqrt{-g}\sum_{j=1}^{9}  
\left( \sum_{i=9}^{11}c_i A^{\tau\sigma}_{ij} + 
       \sum_{k=1}^{3}b_k A^{\sigma\tau}_{kj}  \right) 
J^{\tau\sigma}_j = 0
\end{eqnarray}
Here the result of the variations is expressed as linear combinations 
of the curvature invariants denoted by $J^{\tau\sigma}_j$:
\begin{eqnarray} \label{defJts}
J^{\tau\sigma}_1 &=& \tau^{\mu \nu}\,\nabla_\mu\nabla_\nu\square\sigma \CR
J^{\tau\sigma}_2 &=& R\, \tau^{\mu \nu}\,\nabla_\mu\nabla_\nu\sigma \CR
J^{\tau\sigma}_3 &=&\tau^{\mu \nu}\, \nabla_\mu R\, \nabla_\nu\sigma \CR
J^{\tau\sigma}_4 &=& \tau^{\mu \nu}\, \nabla_\mu \nabla_\nu R\, \sigma \CR
J^{\tau\sigma}_5 &=& \tau\,  R^2\,\sigma \CR
J^{\tau\sigma}_6 &=&  \tau\,\,  \square^2\sigma \CR
J^{\tau\sigma}_7 &=&  \tau\, R \,\square\sigma \CR
J^{\tau\sigma}_8 &=&  \tau \, g^{\mu \nu}\, \nabla_\mu R\,\,\nabla_\nu\sigma \CR
J^{\tau\sigma}_9 &=& \tau \,\square R\, \sigma 
\end{eqnarray}
The coefficients in the result of the $\delta_\tau$ variation in \refb{e18} are
\begin{equation}\label{explAts}
	A^{\tau\sigma}_{ij} = 
\left(
\begin{array}{ccccccccc}
 0 & 0 & 0 & 0 & -2 (x-6) & 0 & 8 & 0 & 0 \\
 -12 & -12 & -36 & 6 (x-6) & 0 & -2 & -2 & 2 & x-6 \\
 24 & 24 & 72 & -12 (x-6) & 2 (x-6) & 12 & 4 & 12 & -6 (x-6)
\end{array}
\right)
\end{equation}
and the coefficients in the $\delta_\sigma$ variations are
\begin{equation} \label{explAst}
	A^{\sigma\tau}_{kj} = \left(
\begin{array}{ccccccccc}
 2 & 2 & 6 & 6-x & 0 & 0 & 0 & -1 & 0 \\
 0 & 0 & 0 & 0 & 6-x & 0 & 4 & 0 & 0 \\
 0 & 0 & 0 & 0 & 0 & 2 & 2 & 4 & 6-x
\end{array}
\right)
\end{equation}
Thus, we have obtained a system of 9 equations, which we use to express 
$b_1$, $b_{2}$, $b_{3}$ in terms of $c_9$, $c_{10}$, $c_{11}$
\begin{eqnarray}
b_1&=& 6 c_{10}-12 c_{11}\CR 
b_2&=& 2 c_{11}-2 c_9\CR 
b_3&=& c_{10}-6 c_{11}
\end{eqnarray}

Since $\Delta_\tau$ does not depend on $B_{\mu \nu\la\rho}$, 
the consistency condition \refb{condtt} is satisfied trivially.

In summary, using conditions \refb{condss} and \refb{condst}, the form 
of the anomalies is reduced to
\begin{eqnarray}
	\Delta_\sigma &=&  
\int d^2x \sqrt{-g}\sum_{j=9}^{11}\sum_{i=1}^{12}  c_j\, M^\sigma_{ji}\, 
I_i \CR \Delta_\tau &=& 
\int d^2x \sqrt{-g} \sum_{j=9}^{11}\sum_{k=1}^{3} c_j \,M^\tau_{jk}\, K_k 
\end{eqnarray}
where
\begin{equation}\label{ms0}
M^\sigma_{ji}=
\left(
\begin{array}{cccccccccc}
 0 & 0 & 0 & 0 & 0 & \frac{6-x}{4} & 0 & 1 & 0 & 0 \\
 0 & -2 & -2 & x-6 & 0 & 0 & -6 & 0 & 1 & 0 \\
 0 & 4 & 4 & -2 (x-6) & 3-\frac{x}{2} & \frac{x-6}{4} & 12 & 0 & 0 & 1
\end{array}
\right)
\end{equation}
and
\begin{equation}\label{mt0}
M^\tau_{jk}=
\left(
\begin{array}{ccc}
 0 & -2 & 0 \\
 6 & 0 & 1 \\
 -12 & 2 & -6
\end{array}
\right)
\end{equation}

Now, we check whether the 
anomalies  $\Delta_\sigma$ and $\Delta_\tau$
are trivial. The most general counterterm $C$ is a linear combination
\begin{equation}\label{explC}
	C = \int d^2x \sqrt{-g} \sum_{j=5}^{7} d_j\, C_j
\end{equation}
of the following curvature invariants
\begin{eqnarray}	\label{defC}
	C_{5} &=& B^{\mu \nu}\,   \nabla_\mu\nabla_\nu R \CR
	C_{6} &=& B\, \square R \CR
	C_{7} &=& B\, R^2
\end{eqnarray}
These are the only possible terms if we take into account partial 
integrations. Variations of $\delta_\sigma$ and $\delta_\tau$ of $C$ 
can be expressed as linear combinations of terms $I_i$ and $K_k$ 
respectively
\begin{eqnarray}
\delta_\sigma C &=&  
 \int d^2x \sqrt{-g}\sum_{l=5}^{7}\sum_{i=1}^{12} d_l\, A'_{li} \,I_i  \CR
\delta_\tau C &=&  
\int d^2x \sqrt{-g}\sum_{l=5}^{7}\sum_{k=1}^{3} d_l \,A''_{lk}\, K_k
\end{eqnarray}
with coefficients given by
\begin{equation}\label{msc}
		A'_{li} = 
\left(
\begin{array}{cccccccccccc}
0 & -2 & -2 & x-6 & 0 & 0 & -6 & 0 & 1 & 0  \\
0 & 0 & 0 & 0 & x-6 & 0 & 0 & -2 & -4 & -2  \\
0 & 0 & 0 & 0 & 0 & x-6 & 0 & -4 & 0 & 0  
\end{array}
\right)
\end{equation}
and
\begin{equation}\label{mtc}
	A''_{lk} =
\left(
\begin{array}{ccc}
 6 & 0 & 1 \\
 0 & 0 & 8 \\
 0 & 8 & 0
\end{array}
\right)
\end{equation}
If we take 
\begin{eqnarray}
d_5&=& c_{10}-2 c_{11}  \CR
d_6&=& -\frac{c_{11}}{2}  \CR
d_7&=& \frac{c_{11}}{4}-\frac{c_9}{4}
\end{eqnarray}
both triviality conditions, \refb{condtriv1} and \refb{condtriv2}, 
are satisfied.

Our conclusion is therefore that not only the trace anomalies found in 
\cite{IMU4} are trivial, but that there cannot be any anomaly 
whatsoever in $J^{(4)\mu}{}_{\mu\la\rho}$.

\section{Conclusion}

In this paper we have applied the trace anomaly method to the calculation
of moments of Hawking radiation. We have shown that, as suggested 
in \cite{IMU4} they 
can be in fact explained as the fluxes of a $W_\infty$ algebra of chiral 
currents, which we have constructed out of two chiral scalar field.
The non--trivial flux of these currents is generated by their response
under a conformal transformation (generalized Schwarzian derivative).
Then we have constructed the covariant and Minkowski version of these 
currents and verified that up to order 6 they are not plagued by any 
trace anomaly, except for $s=2$, i.e. for the energy momentum tensor. 
At this point we have set out to prove that in fact
{\it there cannot exist} any trace anomaly for higher spin currents.
We have succeeded in doing so for the fourth order current and we believe
this is true also for higher order ones\footnote{So it is not very 
appropriate to use the term "trace anomaly method". We should rather
use the term "Schwarzian derivative method"}.

The results of this paper are limited to two dimensions. We do not know
whether they actually extend to four dimensions. The method of 
diffeomorphism anomaly to calculate the Hawking radiation, 
\cite{Robinson}, seem to be more general than the trace 
anomaly method adopted here. It would therefore be very interesting 
to investigate the use of the latter in order to calculate
the higher moments of the Hawking radiation with the same criteria we have 
used in this paper.

\acknowledgments

We would like to thank Silvio Pallua and Predrag Dominis Prester for helpful 
discussions. M.C. would like to thank SISSA for hospitality and CEI and
INFN, Sezione di Trieste, for financial support.

\appendix
\section{Appendix}

Here we write down the order 5 and 6 terms corresponding to \refb{Jn,m}:
\begin{eqnarray}\label{J5}
 J^{(1,4)}_{uuuuu} &=& -3 J^{(1,1)}_{uu} \Gamma ^3-9 J^{(1,2)}_{uuu} \Gamma ^2-4 T J^{(1,1)}_{uu} \Gamma 
-6 J^{(1,3)}_{uuuu} \Gamma +\frac{T \hbar  \left(\partial _uT\right)}{10 }+
\frac{\hbar  \left(\partial _u^3T\right)}{30}\CR
&&+j^{(1,4)}_{uuuuu}-
\left(\partial _uT\right) J^{(1,1)}_{uu}-4 T J^{(1,2)}_{uuu} \CR
 J^{(2,3)}_{uuuuu} &=& -\frac{3}{2} J^{(1,1)}_{uu} \Gamma ^3-3 J^{(1,2)}_{uuu} \Gamma ^2-
\frac{3}{2} J^{(2,1)}_{uuu} \Gamma ^2-T J^{(1,1)}_{uu} \Gamma -J^{(1,3)}_{uuuu} \Gamma 
-3 J^{(2,2)}_{uuuu} \Gamma \CR 
&&-\frac{T \hbar  \left(\partial _uT\right)}{30}
+\frac{\hbar  \left(\partial _u^3T\right)}{60 }+j^{(2,3)}_{uuuuu}-T J^{(2,1)}_{uuu} \CR
 J^{(3,2)}_{uuuuu} &=& -\frac{3}{2} J^{(1,1)}_{uu} \Gamma ^3-\frac{3}{2} J^{(1,2)}_{uuu} \Gamma ^2
-3 J^{(2,1)}_{uuu} \Gamma ^2-T J^{(1,1)}_{uu} \Gamma -J^{(3,1)}_{uuuu} \Gamma -3 J^{(2,2)}_{uuuu} \Gamma 
-\frac{T \hbar  \left(\partial _uT\right)}{30 }\CR
&&+\frac{\hbar  \left(\partial _u^3T\right)}{60 }+j^{(3,2)}_{uuuuu}-T J^{(1,2)}_{uuu} \CR
 J^{(4,1)}_{uuuuu} &=& -3 J^{(1,1)}_{uu} \Gamma ^3-9 J^{(2,1)}_{uuu} \Gamma ^2-4 T J^{(1,1)}_{uu} \Gamma 
-6 J^{(3,1)}_{uuuu} \Gamma +\frac{T \hbar  \left(\partial _uT\right)}{10 }
+\frac{\hbar  \left(\partial _u^3T\right)}{30 }\CR
&&+ j^{(4,1)}_{uuuuu}-\left(\partial _uT\right) J^{(1,1)}_{uu}-4 T J^{(2,1)}_{uuu}
\end{eqnarray}
and
\begin{eqnarray}\label{J6}
	J^{(1,5)}_{uuuuuu} &=& j^{(1,5)}_{uuuuuu}
 +\hbar  \left(\frac{2 T^3}{63}+\frac{5}{42} 
\left(\partial _u^2T\right) T+\frac{17}{168} 
\left(\partial _uT\right)^2+\frac{1}{42} 
\left(\partial _u^4T\right)\right)
 \CR &&
 -\frac{15}{2} J^{(1,1)}_{uu} \Gamma ^4-30 J^{(1,2)}_{uuu} 
\Gamma ^3-15 T J^{(1,1)}_{uu} \Gamma ^2-30 J^{(1,3)}_{uuuu} 
\Gamma ^2-5 \left(\partial _uT\right) J^{(1,1)}_{uu} \Gamma 
-30 T J^{(1,2)}_{uuu} \Gamma 
 \CR &&
 -4 T^2 J^{(1,1)}_{uu}-\left(\partial _u^2T\right) J^{(1,1)}_{uu}-5 
\left(\partial _uT\right) J^{(1,2)}_{uuu}-10 T J^{(1,3)}_{uuuu} -10 J^{(1,4)}_{uuuuu} \Gamma \CR
 J^{(2,4)}_{uuuuuu} &=& j^{(2,4)}_{uuuuuu}
 +\hbar  \left(-\frac{2 T^3}{63}-\frac{2}{105} 
\left(\partial _u^2T\right) T-\frac{1}{840} 
\left(\partial _uT\right)^2+\frac{1}{105} 
\left(\partial _u^4T\right)\right)
 \CR &&
 -3 J^{(1,1)}_{uu} \Gamma ^4-9 J^{(1,2)}_{uuu} \Gamma ^3-3 J^{(2,1)}_{uuu} 
\Gamma ^3-4 T J^{(1,1)}_{uu} \Gamma ^2-6 J^{(1,3)}_{uuuu} \Gamma ^2-9 J^{(2,2)}_{uuuu} 
\Gamma ^2-\left(\partial _uT\right) J^{(1,1)}_{uu} \Gamma 
 \CR &&
 -4 T J^{(2,1)}_{uuu} \Gamma -J^{(1,4)}_{uuuuu} \Gamma -6 J^{(2,3)}_{uuuuu} \Gamma 
-\left(\partial _uT\right) J^{(2,1)}_{uuu}-4 T J^{(2,2)}_{uuuu} -4 T J^{(1,2)}_{uuu} \Gamma \CR
 J^{(3,3)}_{uuuuuu} &=&  j^{(3,3)}_{uuuuuu}
 +\hbar  \left(\frac{2 T^3}{63}-\frac{1}{70} 
\left(\partial _u^2T\right) T-\frac{9}{280} 
\left(\partial _uT\right)^2+\frac{1}{140} 
\left(\partial _u^4T\right)\right)
\CR &&
 -\frac{9}{4} J^{(1,1)}_{uu} \Gamma ^4-\frac{9}{2} J^{(1,2)}_{uuu} 
\Gamma ^3-\frac{9}{2} J^{(2,1)}_{uuu} \Gamma ^3-3 T J^{(1,1)}_{uu} 
\Gamma ^2-\frac{3}{2} J^{(1,3)}_{uuuu} \Gamma ^2-
\frac{3}{2} J^{(3,1)}_{uuuu} \Gamma ^2-9 J^{(2,2)}_{uuuu} \Gamma ^2
 \CR &&
 -3 T J^{(2,1)}_{uuu} \Gamma -3 J^{(2,3)}_{uuuuu} \Gamma -3 J^{(3,2)}_{uuuuu} 
\Gamma -T^2 J^{(1,1)}_{uu}-T J^{(1,3)}_{uuuu}-T J^{(3,1)}_{uuuu} -3 T J^{(1,2)}_{uuu} \Gamma\CR
 J^{(4,2)}_{uuuuuu} &=& j^{(4,2)}_{uuuuuu}
 +\hbar  \left(-\frac{2 T^3}{63}-\frac{2}{105} 
\left(\partial _u^2T\right) T-\frac{1}{840} \left(\partial _uT\right)^2
+\frac{1}{105} \left(\partial _u^4T\right)\right)
 \CR &&
 -3 J^{(1,1)}_{uu} \Gamma ^4-3 J^{(1,2)}_{uuu} \Gamma ^3-9 J^{(2,1)}_{uuu} \Gamma ^3
-4 T J^{(1,1)}_{uu} \Gamma ^2-6 J^{(3,1)}_{uuuu} \Gamma ^2-9 J^{(2,2)}_{uuuu} \Gamma ^2
-\left(\partial _uT\right) J^{(1,1)}_{uu} \Gamma 
 \CR &&
 -4 T J^{(2,1)}_{uuu} \Gamma -J^{(4,1)}_{uuuuu} \Gamma -6 J^{(3,2)}_{uuuuu} \Gamma 
-\left(\partial _uT\right) J^{(1,2)}_{uuu}-4 T J^{(2,2)}_{uuuu} -4 T J^{(1,2)}_{uuu} \Gamma \CR
 J^{(5,1)}_{uuuuuu} &=& j^{(5,1)}_{uuuuuu}
 +\hbar  \left(\frac{2 T^3}{63}+\frac{5}{42} 
\left(\partial _u^2T\right) T
+\frac{17}{168} \left(\partial _uT\right)^2+\frac{1}{42} 
\left(\partial _u^4T\right)\right)
\CR &&
 -\frac{15}{2} J^{(1,1)}_{uu} \Gamma ^4-30 J^{(2,1)}_{uuu} \Gamma ^3-15 T J^{(1,1)}_{uu} 
\Gamma ^2-30 J^{(3,1)}_{uuuu} \Gamma ^2-5 \left(\partial _uT\right) J^{(1,1)}_{uu} 
\Gamma -30 T J^{(2,1)}_{uuu} \Gamma 
 \CR && 
 -4 T^2 J^{(1,1)}_{uu}-\left(\partial _u^2T\right) J^{(1,1)}_{uu}-5 
\left(\partial _uT\right) J^{(2,1)}_{uuu}-10 T J^{(3,1)}_{uuuu}-10 J^{(4,1)}_{uuuuu} \Gamma
\end{eqnarray}

\end{document}